\documentclass[aps,prd,showpacs,floatfix,nofootinbib,twocolumn,10pt]{revtex4}

\usepackage{color,amsmath,amssymb,graphicx,latexsym,subfigure}

\begin{document}

\title{Gamma-rays From Warm WIMP Dark Matter Annihilation}

\author{Qiang Yuan$^1$, Yixian Cao$^2$, Jie Liu$^3$, Peng-Fei Yin$^1$,
Liang Gao$^2$, Xiao-Jun Bi$^1$ and Xinmin Zhang$^3$}

\affiliation{$^1$Key Laboratory of Particle Astrophysics, Institute of
  High Energy Physics, Chinese Academy of Sciences,
  Beijing 100049, P.R.China\\
  $^2$Partner Group of the Max Planck Institute for Astrophysics,
  National Astronomical Observatories, Chinese Academy of Sciences,
  Beijing, 100012, P.R. China\\
  $^3$Division of Theoretical Physics, Institute of High Energy
  Physics, Chinese Academy of Science, Beijing 100049, P.R.China}

\date{\today}

\begin{abstract}

The weakly interacting massive particle (WIMP) often serves as a candidate
for the cold dark matter, however when produced non-thermally it could
behave like warm dark matter. In this paper we study the properties of
the $\gamma$-ray emission from annihilation of WIMP dark matter in the halo
of our own Milky-Way Galaxy with high resolution $N$-body simulations of
a Milky-Way like dark matter halo, assuming different nature of WIMPs.
Due to the large free-streaming length in the scenario of warm WIMPs,
the substructure contend of the dark matter halo is significantly different
from that of the cold WIMP counterpart, resulting in distinct predictions
of the $\gamma$-ray signals from the dark matter annihilation. We illustrate
these by comparing the predicted $\gamma$-ray signals from the warm
WIMP annihilation to that of cold WIMPs. Pronounced differences from the
subhalo skymap and statistical properties between two WIMP models are
demonstrated. Due to the potentially enhanced cross section of the
non-thermal production mechanism in warm WIMP scenario, the Galactic
center might be prior for the indirect detection of warm WIMPs to dwarf
galaxies, which might be different from the cold dark matter scenario.
As a specific example we consider the non-thermally produced neutralino
of supersymmetric model and discuss the detectability of warm WIMPs with
Fermi $\gamma$-ray telescope.

\end{abstract}

\pacs{95.35.+d,95.85.Pw}

\maketitle

\section{Introduction}

The so-called dark matter (DM), discovered $\sim80$ years ago in the
astronomical observations, is still one of the biggest mysteries in
the fields of physics, astronomy and cosmology. To understand the
nature of DM particles is a big challenge of the community. There
are several ways being proposed to detect the WIMP DM particles (see e.g.,
\cite{2005PhR...405..279B}), among which the {\it indirect search}
through the cosmic ray (CR) particles is the most active one in recent
years due to the operation of several new generation satellites, such
as PAMELA, Fermi and AMS02. In many kinds of CR particles, the
anti-particles, $\gamma$-rays and neutrinos are good probes to search
for DM signals. Especially, $\gamma$-rays are widely discussed, due to
the simple propagation and the high sensitivity detections from both
spatial and ground-based telescopes. The constraints on the DM parameters
become stronger and stronger in recent years thanks to the Fermi
$\gamma$-ray observations \cite{2010ApJ...712..147A,2010JCAP...05..025A,
2010JCAP...04..014A,2011PhRvL.107x1302A,2011PhRvL.107x1303G}.

One of the key problems in the study of the $\gamma$-ray emission from the
WIMP DM annihilation is the density distribution of DM. It is observationally
very difficult to determine the density distribution of DM, especailly at
small scales. Currently the postulated best knowledge about the DM density
distribution comes from the numerical N-body simulations (e.g.,
\cite{2008Natur.456...73S,2008MNRAS.391.1685S,2008Natur.454..735D}).

The initial matter power spectrum which describes cosmic density
perturbation depends on the particle nature of DM. For the cold DM (CDM),
the particle velocity when decoupling is negligible and the corresponding
free-streaming length is very short. The small free-streaming
length enables structures down to very small scales to form.

The CDM scenario has been shown to be in good agreement with the
observations of the cosmological large scale structures. However, it has
been a long time problem of the CDM scenario that the expected structures
are inconsistent with observations at sub-galactic scale
(e.g., \cite{1994Natur.370..629M,1995ApJ...447L..25B,1999ApJ...524L..19M,
1999ApJ...522...82K,2004ApJ...617.1059R}). One possible solution of this
problem is the warm DM (WDM) scenario (\cite{2000ApJ...542..622C,
2001ApJ...551..608S,2001ApJ...556...93B}, or a recent review
\cite{2012EAS....53...97T}). In general, with a thermal distribution, the
particle mass of the WDM should be as light as $\sim$keV. After decoupling
the velocity of WDM can be fast enough to introduce a large free-streaming
scale below which the structures are smoothed out. Thus the formation of
small scale structures in the WDM scenario can be suppressed.

If the DM is finally proven to be warm, the impact on the detection of DM
particles is fatal, because most of these experiments aim to search for
the weakly interacting massive particles (WIMPs) which are traditionally
cold. For the canonical WIMPs, when produced thermally in the early
Universe, the velocity is non-relativistic after decoupling and they behave
like CDM. Alternatively the WIMPs if produced non-thermally, it can be warm
\cite{1999JHEP...12..003J,2000astro.ph..8451H,2001PhRvL..86..954L,
2002PhRvD..66h3501F}. In Ref.\cite{2001PhRvL..86..954L} the authors showed
explicitly that the power spectrum of this non-thermally produced WIMPs has
a clear suppression at small scales. The non-thermally produced WIMP
scenario will have some interesting properties for the {\it indirect
search} of DM, because 1) compared with the light WDM the mass of the
non-thermal WIMPs lies within the range of most high energy CR detectors,
and 2) in contrast with the thermally produced WIMPs, the annihilation
cross section of non-thermal WIMPs can be larger due to the lack of
direct constraints from the relic density. We will discuss the possible
$\gamma$-ray signatures from such non-thermal WIMP DM annihilation in
this paper.

In this paper we focus on predicted DM annihilation signals from the Milky
Way halo and its substructures based upon high resolution simulations of
WDM in \cite{2012MNRAS.420.2318L}.

This paper is organized as follows. In Sec. II we briefly introduce the
picture of the non-thermally produced warm WIMPs. In Sec. III we describe
the numerical simulations used in this work and the DM density distributions
for the smooth halo and subhalos according to the simulations. The
signatures of $\gamma$-ray signals and detectability analysis are discussed
in Sec. IV. Finally Sec. V is the conclusion.

\section{Non-thermally produced warm WIMPs}

The DM particles can be non-thermally produced by the decays of topological
defects such as cosmic string \cite{1999JHEP...12..003J,2001PhRvL..86..954L,
2009PhRvD..79h3532C,2009PhRvD..80j3502B}. For example, we consider a model
with an extra U(1) gauge symmetry which is broken by the vacuum expectation
value $\eta$ of a scalar field $S$ \cite{1999JHEP...12..003J}. Cosmic
strings will be formed during the symmetry breaking phase
transition taking place at the temperature of $T_c \sim \eta$. After the
transition, the infinite long string network coarsens, and more closed
string loops form from the reconnection of the long strings. The tension
of the cosmic string is determined by $\mu \sim \eta^2 $. Cosmic string
loops lose their energy dominantly through gravitational radiation. When
the radius of a loop becomes to the order of the string width, the loop
will self-annihilate into its constituent field, such as scalar boson $S$.
The DM particle $\chi$ can be produced by the decay of these heavy particles.

When the temperature of DM is higher than the freeze-out temperature 
$T_\chi$($\sim$ O(GeV)), DM particles produced by cosmic string loops 
still keep in chemical equilibrium with standard model particles. Only 
the DM particles produced below $T_\chi$ will contribute to the non-thermal 
DM relic density $\Omega_{\rm NT}$. It is found that the non-thermal DM 
is mostly contributed by the loops decaying at $T_\chi$ (Eq. (A2)). 
Therefore the DM production process does not affect the big bang 
nucleosynthesis (BBN) results. Through adjusting the model parameters, 
the relic density of DM can also be naturally explained (for more details, 
see Appendix A).

In such scenario, DM particle $\chi$ may carry large initial momentum $p_c$
due to the decay of the heavy particle. $p_c$ can be written as $p_c = \alpha
T_c$ where $\alpha$ is a numerical factor determined by detailed model.
Here we define a typical variable $r_c = a(t) p(t) /m_\chi $ which is
a constant during the cosmic evolution \cite{2001PhRvL..86..954L}.
If we choose the cosmic scale factor at the present time $a(t_0)=1$,
$r_c$ can be understood as today's velocity of the DM particles if no
structure formation. The comoving free-streaming scale $R_f$ is given
by \cite{2001PhRvL..86..954L}
\begin{eqnarray}
R_f &=& \int^{t_{\rm EQ}}_{t_i} \frac{v(t')}{a(t')} {\rm d}t' \nonumber \\
&\sim & 2 r_c t_{\rm EQ} (1+z_{\rm EQ})^2 \ln \left[ \sqrt{1+
\frac{1}{r_c^2 (1+z_{\rm EQ})^2}}\right. \nonumber \\
&+& \left.\frac{1}{r_c (1+z_{\rm EQ})} \right],
\end{eqnarray}
where ``EQ'' denotes the radiation-matter equality.

\begin{figure}[!htb]
\centering
\includegraphics[width=0.9\columnwidth]{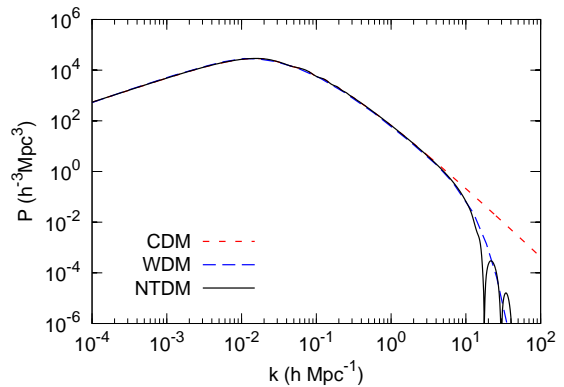}
\caption{Linear matter power spectra of CDM (red, short-dashed), canonical
light WDM (blue, long-dashed) and non-thermal warm WIMP (black, solid).
\label{fig:pk}}
\end{figure}

The free-streaming of DM particles will imprint on the late time structure
formation. This effect can be simply seen by the matter power spectrum of
DM. We use a modified version of CAMB\footnote{http://camb.info}
\cite{2000ApJ...538..473L} to calculate the matter power spectrum of the
non-thermally produced DM scenario, shown in Fig. \ref{fig:pk}.
Here we adopt $r_c=10^{-7}$. Note the mass of the non-thermal DM does
not explicitly affect the calculation of the power spectrum because its
effect can be cancelled by the initial momentum (see the definition of
$r_c$). For comparison the power spectra for CDM and the canonical light
WDM are also shown. The power spectrum of the canonical WDM corresponds to
a sterile neutrino with mass $\sim 2$ keV, which is also the input power
spectrum of the N-body simulation (see below Sec. III). We can see that a
clear suppression of the power at small scales appears both for the light
WDM and the warm WIMP scenarios. The free-streaming property makes the
non-thermal DM behave similarly with WDM. Due to the similarity of the
input power spectra of the light WDM and non-thermal warm WIMPs, we use
the simulation results for the light WDM in the following discussion of
the {\it indirect detection} of warm WIMPs.

\section{Numerical simulation results}

In this section, we describe briefly numerical simulations used in this
work and present the properties of the DM distribution based on the
numerical simulations. The simulations used in this study are two matched
ultra-high resolution simulations of a Milky sized DM halos run with
different nature of DM models but with otherwise same numerical setup
as well as cosmological parameters. For the CDM simulation, we use
``Aq-A-2'', from the {\it Aquarius} Project \cite{2008MNRAS.391.1685S}.
In order to facilitate comparison of DM annihilation emission from cold
and warm DM models, for the same halo, we further performed a high
resolution simulation assuming WDM model by using the same phase in the
initial density field as that of the ``Aq-A-2'' simulation but a different
matter spectrum matching a particular WDM model. In the numerical calculation
of this paper, we adopt a $2$ kev sterile neutrino \cite{2009PhRvL.102t1304B}
as our WDM model which lies within bound of {\it Ly$\alpha$}
constraint\cite{2009PhRvL.102t1304B}. The chosen WDM introduce a
cutoff emerging at a wavenumber $k\sim10h$ Mpc$^{-1}$ in the initial
matter power spectrum, below which the power spectrum is well
consistent with that of CDM \cite{2012MNRAS.420.2318L}. In the
scenario of non-thermal WIMPs, such as a heavy particle $S$ decaying
into two WIMPs $\chi$, for $\chi$ around $100$ GeV, it requires the mass
of particle $S$ around $10^8$ GeV \cite{2001PhRvL..86..954L}. In our
simulation, the mass of the ``particle'' is $1.37\times10^4$
M$_{\odot}$, and the number of ``particles'' is larger than $100$
million within $r_{200}$, the radius inside which the mean DM density is $200$
times of the critical density. Therefore the lowest mass subhalos resolved
in our simulation is $3\times10^5$ M$_{\odot}$ if requiring more than $20$
``particles'' for a subhalo. The total mass within $r_{200}$ of the halo
is about $1.8\times10^{12}$ M$_{\odot}$. See Table 1 of Ref.
\cite{2012MNRAS.420.2318L} for the basic information of the simulations.
For a more detailed description of our simulation, please refer to Ref.
\cite{2012MNRAS.420.2318L}.

\subsection{Smooth halo}

The density profile of the smooth component of the simulated halo of the
CDM simulation was analyzed in Ref. \cite{2010MNRAS.402...21N}. It was
shown that the smooth halo density profile can be well fitted with an
Einasto profile \cite{Einasto:1965}
\begin{equation}
\rho(r)=\rho_{-2}\exp\left[-\frac{2}{\alpha}\left(\left(\frac{r}
{r_{-2}}\right)^{\alpha}-1\right)\right],
\label{rho_ein}
\end{equation}
where $\rho_{-2}\approx 0.14$ GeV cm$^{-3}$, $r_{-2}\approx 15.7$ kpc
and $\alpha\approx 0.17$ \cite{2010MNRAS.402...21N}. The local density
of DM is then given as $\rho_{\odot}\approx 0.44$ GeV cm$^{-3}$ at
$R_{\odot}=8.5$ kpc. A higher local density compared with the canonical
$0.3$ GeV cm$^{-3}$ was also found in recent studies
\cite{2010JCAP...08..004C,2010A&A...523A..83S,2010PhRvD..82b3531P}.

For the WDM halo the density profile of the smooth halo is essentially the
same as that of CDM down to the numerical resolution limit of our simulation
\cite{Shao-Gao:WDM}. The expectation that a core may appear in the center
of the halo for WDM due to phase space density constraint
\cite{2004PhRvD..69l3521B,2005PhRvL..95r1301C,2005PhRvD..72f3510K} is not
clearly seen in the simulation, this is because the core size of the Milky
Way sized halo is predicted to be smaller than resolution limit of our
simulation and thus is not resolved. It was shown recently that a density
core was indeed observed in WDM simulations, at a scale smaller than $100$
pc for $1-2$ keV WDM and halo mass $10^8-10^{10}$ M$_{\odot}$
\cite{2012arXiv1202.1282M}. For the Milky Way like halo the expected core
will be even smaller, and the halo density profile will be indistinguishable
from that of CDM halo, within the precision of sub-degree of the present
$\gamma$-ray detectors. In this work we adopt the same equation
(\ref{rho_ein}) to describe the density profile of the smooth halo for WDM.

\subsection{Subhalos}

Based on the simulation results, we find $20529$ gravitational bounded
subhalos for CDM simulation and $219$ subhalos for WDM\footnote{Note
for WDM case, the number of subhalos might be over-estimated due to the
numerical fragmentation of filaments \cite{2007MNRAS.380...93W}.}
simulation within the virial radius of the main halo. The minimum mass
of the resolved subhalo is found to be $\sim 3\times10^5$ M$_{\odot}$,
and the maximum mass is about $10^{10}$ M$_{\odot}$.

We define the annihilation luminosity of a subhalo as $L_i= \int
\rho_i^2 {\rm d}V_i$. In the work we adopt the Navarro-Frenk-White
(NFW, \cite{1997ApJ...490..493N}) profile for the subhalos. The
determination of the parameters of the NFW density profile from
the simulated circular velocity profile can be found in the Appendix of
Ref. \cite{1997ApJ...490..493N}. For WDM subhalos we employ a constant
density core with size $r_c\approx 0.03\times\left(\frac{\sigma}
{{\rm km\,s^{-1}}}\right)^{-0.5}$ kpc, where $\sigma$ is the velocity
dispersion of the subhalo \cite{2012arXiv1209.5563S}. Beyond $r_c$ the
density distribution is identical with NFW profile.
The $\gamma$-ray flux from DM annihilation
of this subhalo is then proportional to $L_i/d_i^2$, with $d_i$ the
distance of the subhalo from Earth. To calculate $d_i$ of each subhalo,
a random location of the solar system which is $8.5$ kpc away from the
halo center is chosen.

\begin{figure*}[!htb]
\centering
\includegraphics[width=0.9\columnwidth]{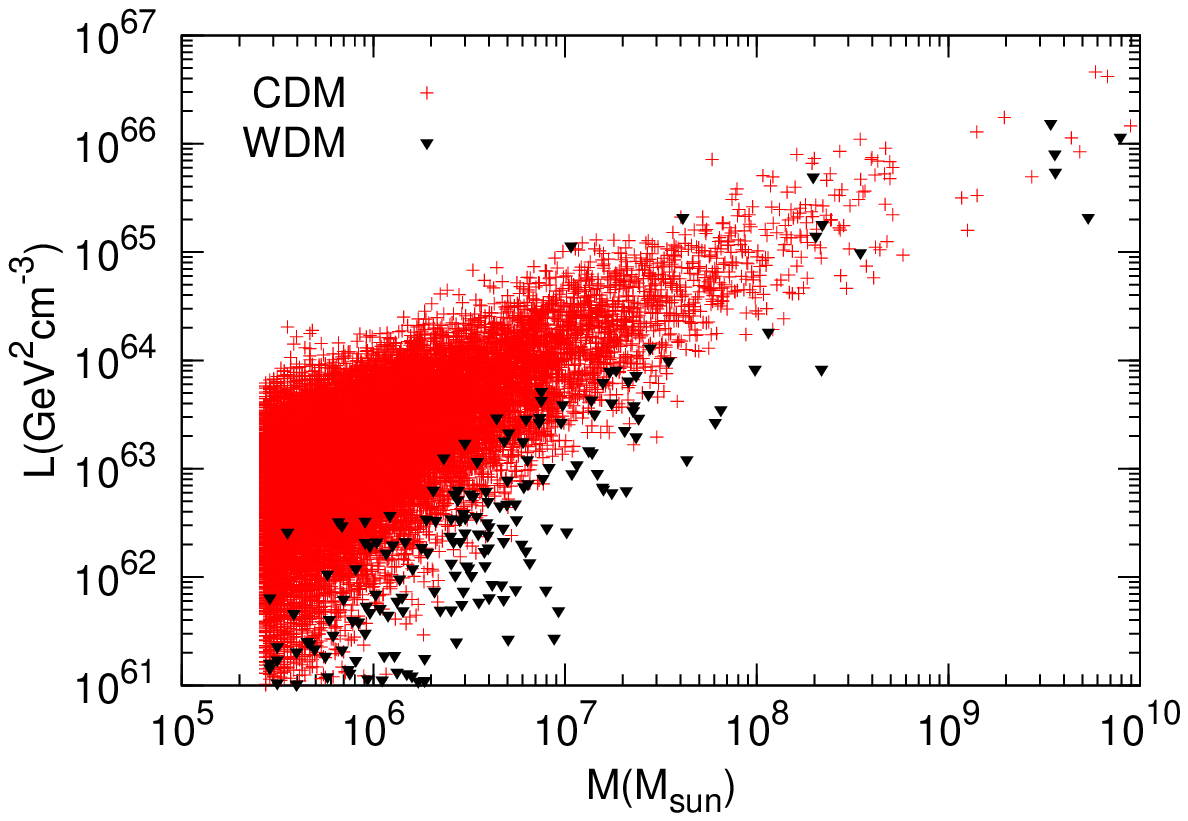}
\includegraphics[width=0.9\columnwidth]{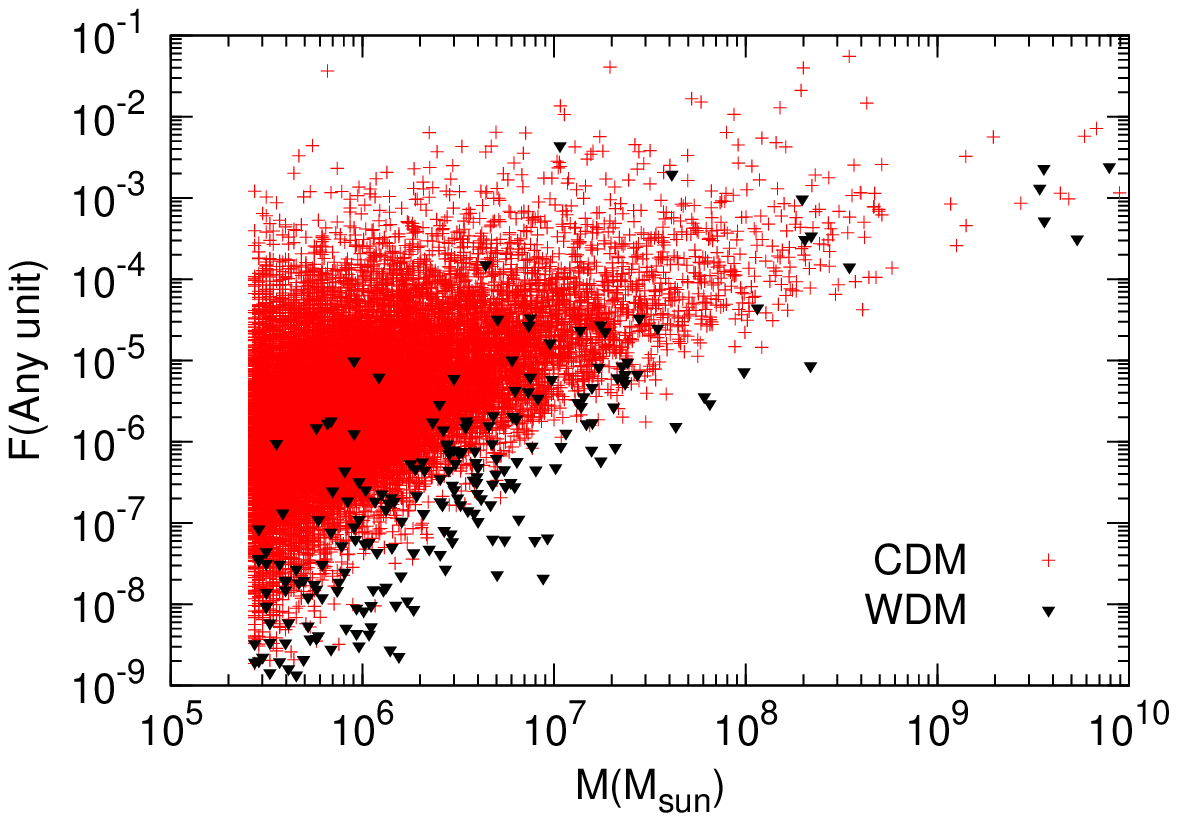}
\caption{Annihilation luminosity ($L$, left) and relative flux ($F$, right)
versus mass of subhalos for CDM and WDM simulations.
\label{fig:lumin_flux}}
\end{figure*}

The mass-luminosity and mass-flux scattering plots of the subhalos are
shown in Fig. \ref{fig:lumin_flux}. From the mass-flux relation we see
that in general, subhalos in CDM case are more brighter than that of
WDM because of relatively lower concentration of subhalo in WDM comparing
to CDM \cite{2012MNRAS.420.2318L}. There are also fewer subhalos of WDM
which can have comparable fluxes as that of CDM. Especially we find the
most massive subhalos are usually not the brightest objects. The subhalos
with masses $10^7-10^9$ M$_{\odot}$ have larger probability to give
high fluxes \cite{2009ChPhC..33..826Y}.

For the CDM case it is expected that there should be a large number of
unresolved substructures below the resolution limit of the simulation,
which can extend to a mass comparable or even lower than that of the
Earth, $10^{-6}$ M$_{\odot}$ \cite{2001PhRvD..64h3507H,
2005Natur.433..389D}. To include the contribution of unresolved subhalos
we have to extrapolate the subhalos to lower mass, according to the
statistical properties of the resolved subhalo distributions
\cite{2008MNRAS.391.1685S,2011MNRAS.410.2309G}. We present the basic
statistical results of the subhalos of CDM and WDM based on the simulations
in the Appendix. For the WDM case, because free streaming length
of the chosen WDM particle is as large as $200$ kpc, the smallest
dark matter halo expected to form in the model is therefore about
$2.5 \times 10^{9}$ M$_{\odot}$ \cite{2001ApJ...556...93B}, corresponding 
to $\sim10^5$ particles in our simulation and hence are well resolved
in our simulation. Thus we believe that our WDM simulation has
resolved all subhalos and theorefore has no unresolved subhalo
compounent. There are some spurious subhalo formed in our
simulation via artificial fragmentation of filaments as noted by
Ref. \cite{2007MNRAS.380...93W}, however we expect that contribution
to the annihilation luminosity due to these spurious subhalos is
small because of their low abundance. We do not consider them in the 
following analysis.

\subsection{$J$-factors}
The $\gamma$-ray flux observed at the Earth from DM annihilation can be
written as
\begin{equation}
\phi_{\gamma}(E_{\gamma},\psi)=\frac{\rho_{\odot}^2R_{\odot}}{4\pi}
\frac{\langle\sigma v\rangle}{2m_{\chi}^2}\frac{{\rm d}N}{{\rm d}
E_{\gamma}}\times J(\psi),
\end{equation}
where $m_\chi$ is the mass of the DM particle, $\langle\sigma v\rangle$
is the annihilation cross section weighted with the velocity of DM
particle, $\frac{{\rm d}N}{{\rm d}E_{\gamma}}$ is the $\gamma$-ray yield
spectrum per annihilation. The dimensionless astrophysical $J$-factor,
related to the DM density profile, is defined as
\begin{equation}
J(\psi)=\frac{1}{\rho_{\odot}^2R_{\odot}}\int_{LOS}\rho^2(l){\rm d}l,
\label{jpsi}
\end{equation}
where $\psi$ is defined as the angle between the observational direction
and the Galactic center direction for observer at the Earth. The integral is
done along the line of sight (LOS). Taking the detectors angular
resolution into account the $J$ factor for a resolved subhalo is
defined as
\begin{equation}
J_{\rm sub}^{i}(\psi)=\frac{1}{\rho_{\odot}^2R_{\odot}}\frac{L_i}{d_i^2}
\times\frac{1}{2\pi\sigma^2}\exp\left[-\frac{(\psi-\psi_i)^2}{2\sigma^2}\right],
\label{jpsi_sub}
\end{equation}
where $L_i$, $d_i$ and $\psi_i$ are the luminosity, distance and central
direction of the $i$th halo. The exponential term on the right hand side
corresponds to a Gaussian smooth with width $\sigma$.

\begin{figure*}[!htb]
\centering
\includegraphics[width=0.9\columnwidth]{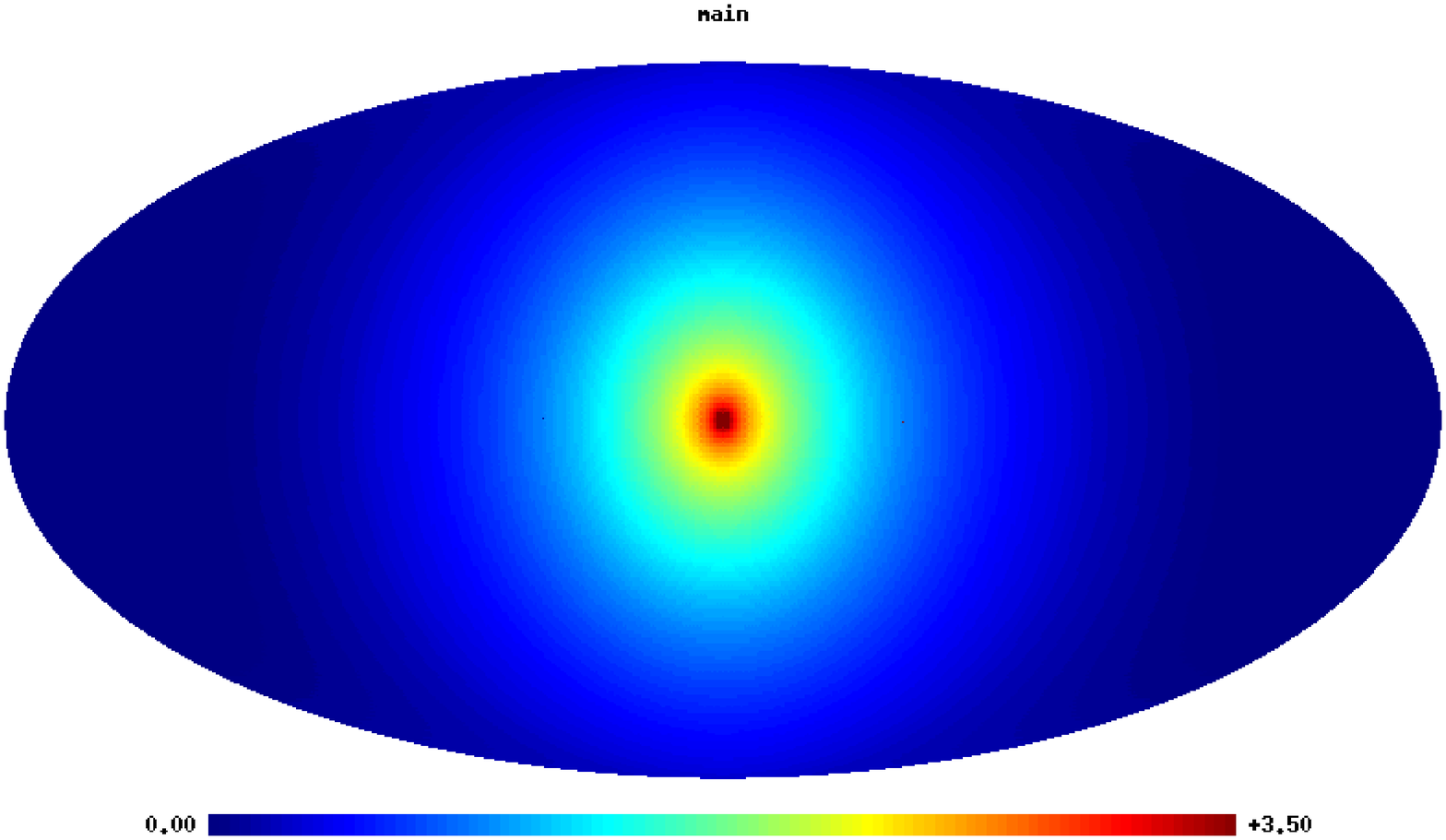}
\includegraphics[width=0.9\columnwidth]{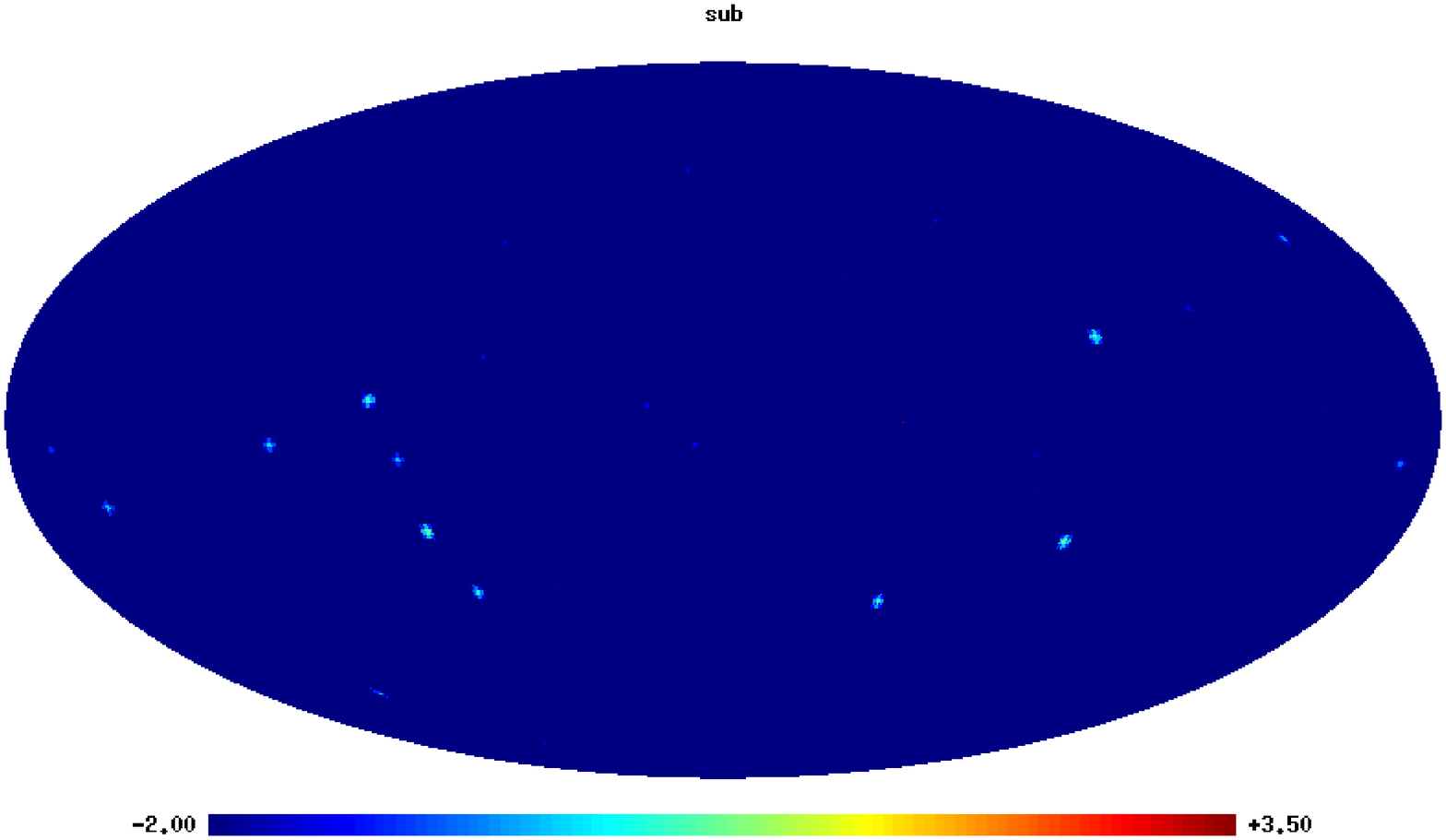}
\includegraphics[width=0.9\columnwidth]{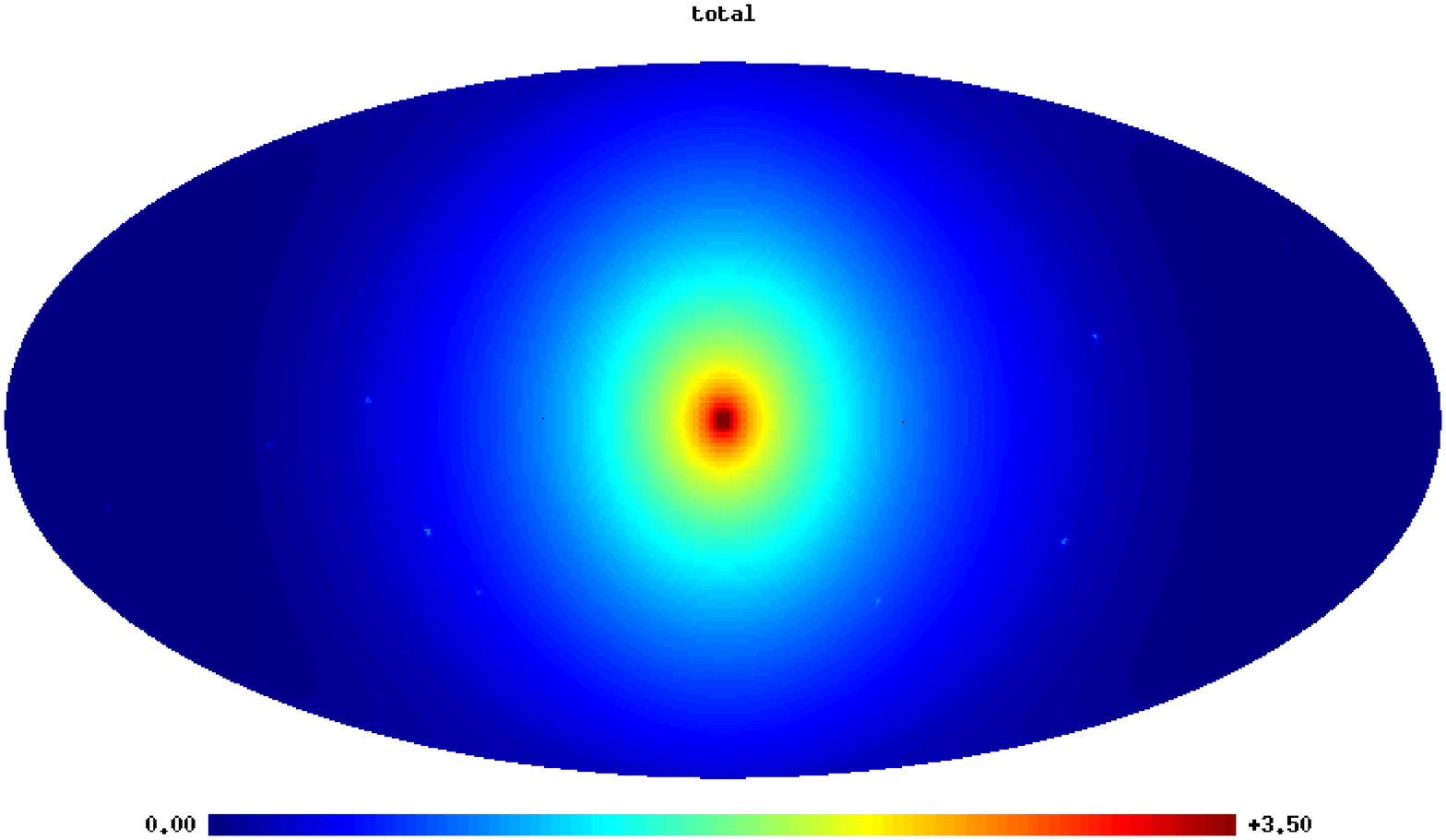}
\caption{Skymaps of the $J$-factors of the main halo (top-left),
resolved subhalos (top-right) and total contribution (bottom) for WDM.
\label{fig:skymap_wdm}}
\end{figure*}

Based on the numerical simulation of WDM we calculate the $J$ factor
of the smooth halo and the subhalos. The skymaps of the $J$ factors
of the smooth halo, resolved subhalos and the total result for WDM are
shown in Fig. \ref{fig:skymap_wdm}. The colorbar shows the value of
$\log(J)$. For resolved subhalos we employ Gaussian smoothing with
$\sigma=0.5^{\circ}$.

The skymaps of the CDM subhalos based on the simulation {\it Aquarius}
has been given in  Ref. \cite{2008Natur.456...73S}. To compare with
the skymap of WDM subhalos given in this work we have also shown
the skymaps of CDM in Appendix (see Fig. \ref{fig:skymap_cdm}).
From those two figures we can clearly see the differences between
the CDM and WDM annihilation signals from subhalos. For CDM, there is
a non-negligible diffuse component from the unresolved subhalos,
especially at the directions far away from the Galactic center. The
number of the potentially visible subhalos above the diffuse component
is much higher for CDM than WDM. There is also difference in the Galactic
center due to expected presence of a core in the WDM scenario. However,
we may over-estimate the size of the core in this work compared with
that found in the simulations \cite{2012arXiv1202.1282M}. The actual
difference may be smaller.

\begin{figure}[!htb]
\centering
\includegraphics[width=0.9\columnwidth]{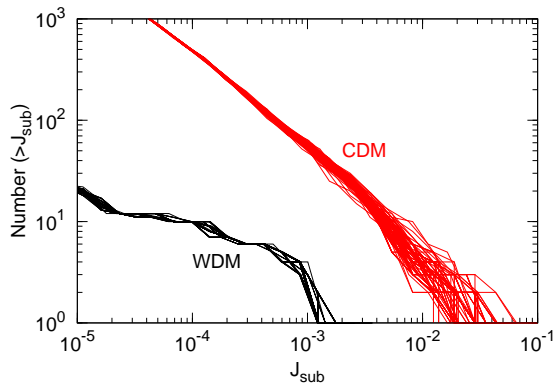}
\caption{Accumulative number versus $J$ of subhalos.
\label{fig:NJfactor}}
\end{figure}

The accumulative subhalo number which represents the subhalos with $J$
factor greater than some value $J_{\rm sub}$ versus $J_{\rm sub}$
is shown in Fig. \ref{fig:NJfactor}. The different lines in each group
represent the random choice of the location of the solar system in the
halo, with distance fixed to be $8.5$ kpc from the center. We can see that
the number distribution of WDM is flatter than that of CDM. This is because
in the CDM case the relative weight of smaller subhalos compared with
larger ones are higher than that in the WDM case. The property presented
in Fig. \ref{fig:NJfactor}, if detectable, is useful to probe the
nature of the DM particles.

\section{Gamma-ray signals}

In this section we study the $\gamma$-ray signals from the warm WIMP
annihilation. We will present the astrophysical $\gamma$-ray background
and the detectability of the $\gamma$-rays from warm WIMP annihilation by
Fermi.

\subsection{Benchmark models of supersymmetric DM}

\begin{table*}[htb]
\centering
\caption{Relevant parameters for the two benchmark models. The unit of
$m_{0}$, $m_{_{H_u}}$, $m_{_{H_d}}$, $m_{1/2}$, $A_0$,
$m_{\tilde{\chi}_1^0}$ is GeV, and of $\langle\sigma v\rangle$
is cm$^3$s$^{-1}$.}
\begin{tabular}[c]{cccccccccc}
\hline\hline
& \;\; $m_{0}$ \;\; & \; $m_{_{H_u}}$ \;& \; $m_{_{H_d}}$ \;& \; $m_{1/2}$ \;& \;$A_0$ \;& $\tan \beta$ & ${\rm sign}(\mu)$ &\; $m_{\tilde{\chi}_1^0}$ \; & $\langle\sigma v \rangle$\\
\hline
Warm WIMP &  1200  & 1300   &  788   & 500   & -1000  & 40   & +   & 211 & $2.70\times 10^{-25}$\\
\hline
Cold WIMP &  1200  & 1300   &  824   & 500   & -1000  & 40   & +   & 211 & $1.38 \times 10^{-26}$ \\
\hline\hline
\end{tabular}
\label{tablebmp}
\end{table*}

For the warm WIMP, the annihilation cross section may be larger than
that of cold WIMP which are constrained by the relic density of
DM. However, considering the constraints from e.g., $\gamma$-rays and
antiprotons, the cross section can not be arbitrarily large. The
constraint from PAMELA antiproton data showed that the allowed boost
factor\footnote{Defined as $\langle\sigma v\rangle/3\times10^{-26}$
cm$^3$s$^{-1}$.} of neutralino-like DM should be less than $10$ for
$O(100)$ GeV DM \cite{2009PhRvL.102g1301D}. The new constraints from
Fermi observations of dwarf galaxies also gave an allowed boost factor
of several for $O(100)$ GeV DM \cite{2011PhRvL.107x1302A}.
Taking the above contraints on the WIMP annihilation cross section
into account, we give two explicit benchmark models to realize the
cold and warm WIMP scenarios in supersymmetric DM models.

In the supersymmetric (SUSY) theory with R-parity conservation, the
lightest neutralino, which is the combination of gaugino and higgsino,
is a well-motivated candidate of DM \cite{1996PhR...267..195J}. In general,
there are four parameter regions to obtain the correct thermal relic density
of neutralino: (1) all the sfermions are light, neutralinos annihilate via
$t$-channel sfermions exchange; (2) neutralinos scatter with sfermions with
nearly mass degeneracy which is so-called ``co-annihilation''; (3)
$\tilde{\chi}_1^0$ has significant component of Higgsino or wino, with the
main annihilation channel to heavy gauge boson or Higgs; (4) neutralinos
annihilate via $s$-channel Higgs resonance with $2m_{\tilde{\chi}_1^0}
\sim m_{A^0}$, or $m_{h^0}$, $m_{H^0}$. In the first region, the light
sfermions are stringently constrained by recent LHC results
\cite{2011PhRvL.107v1804C,2012PhLB..710...67A}. In the ``co-annihilation''
region, the neutralino annihilation cross section is often much smaller
than the ``natural value'' $3\times 10^{-26}$ cm$^3$s$^{-1}$. Thus it is
difficult to observe the products of DM annihilation in indirect
detections. In the third region, neutralino annihilation could produce
large flux of $\gamma$-rays due to cascade decay of gauge boson or Higgs.
However, a significant component of Higgsino or wino in the neutralino
might induce large interaction between DM and nucleon, which is
stringently constrained by recent direct detections, such as XENON100
\cite{2011PhRvL.107m1302A}.

Here we consider two benchmark models in the ``Higgs funnel'' region as
the cold and warm WIMP candidates. The DM annihilation is enhanced by
$s$-channel pseudoscalar Higgs exchange with resonance effect
$m_{A^0}\sim 2m_{\tilde{\chi}_1^0}$, and the main final states of DM
annihilations are $b\bar{b}$\footnote{The potential phenomenology
problem of this region may be the large contribution to rare meson
decay, such as $B_d \to X_s \gamma $ and $B_s \to \mu^+ \mu^-$, due
to light pseudoscalar Higgs and large $tan \beta$. To avoid violating
meson decay observations, some special parameters are needed to suppress
total SUSY contributions from Higss sector and chargino-squark sector.}.
The ATLAS and CMS collaborations have discovered a 125 GeV Higgs-like 
boson \cite{2012PhLB..716....1A,2012arXiv1207.7235C}. Because the Higgs 
in MSSM is lighter than $Z$ boson at the tree level, it requires large stop 
mass parameter or large mixing term to acquire corrected Higgs mass. 
It can be interpreted by some particular parameter configurations. Since
here we employ the benchmark models as illustration and emphasize the
difference between warm and cold WIMPs, we do not consider this issue 
of Higgs mass in this work.

To acquire a moderate $A^0$ mass easily, we consider the ``NUHM'' scenario
\cite{2003NuPhB.652..259E,2005JHEP...07..065B}, in which the Higgs mass
parameters $m_{H_u}$,$m_{H_d}$ at GUT scale are different from other scalar
masses $m_{0}$. The particle spectrum, DM thermal relic density and
annihilation cross section for the bench-marks models are calculated by
SuSpect \cite{2007CoPhC.176..426D} and micrOMEGAs \cite{2007CoPhC.176..367B,
2009CoPhC.180..747B}, and are summarized in Table \ref{tablebmp}. For the 
warm WIMP model adopted here, the thermal relic density of DM is $\Omega 
h^2_{\rm th}\sim4.33\times10^{-3}$, much smaller than the observational 
value $\Omega h^2\sim0.11$. Therefore it must be produced via non-thermal 
mechanism (see Appendix A and \cite{2009PhRvD..80j3502B}). 
Given the particle models of DM, the $\gamma$-ray spectrum from the 
decay and fragmentation of the final state particles is calculated 
using the PYTHIA simulation tool \cite{2006JHEP...05..026S}.

In the benchmark models the DM parameters are $m_{\chi}\approx211$ GeV,
$\langle\sigma v\rangle\approx1.38\times 10^{-26}$ cm$^3$ s$^{-1}$ for
cold WIMP and $2.70\times10^{-25}$ cm$^3$s$^{-1}$ for warm WIMP, and
the annihilation final state is about $86\%\,b\bar{b}+14\%\tau^+\tau^-$.
The cross section for warm WIMP corresponds to a boost factor of $9$,
which is roughly compatible with the present constraints from {\it
indirect detections}. Note these constraints are applicable for neutralino
DM. For other annihilation final states such as the leptons the constraints
might be different, and even larger boost factor could be possible.

\subsection{Astrophysical background}

\begin{figure*}[!htb]
\centering
\includegraphics[width=0.9\columnwidth]{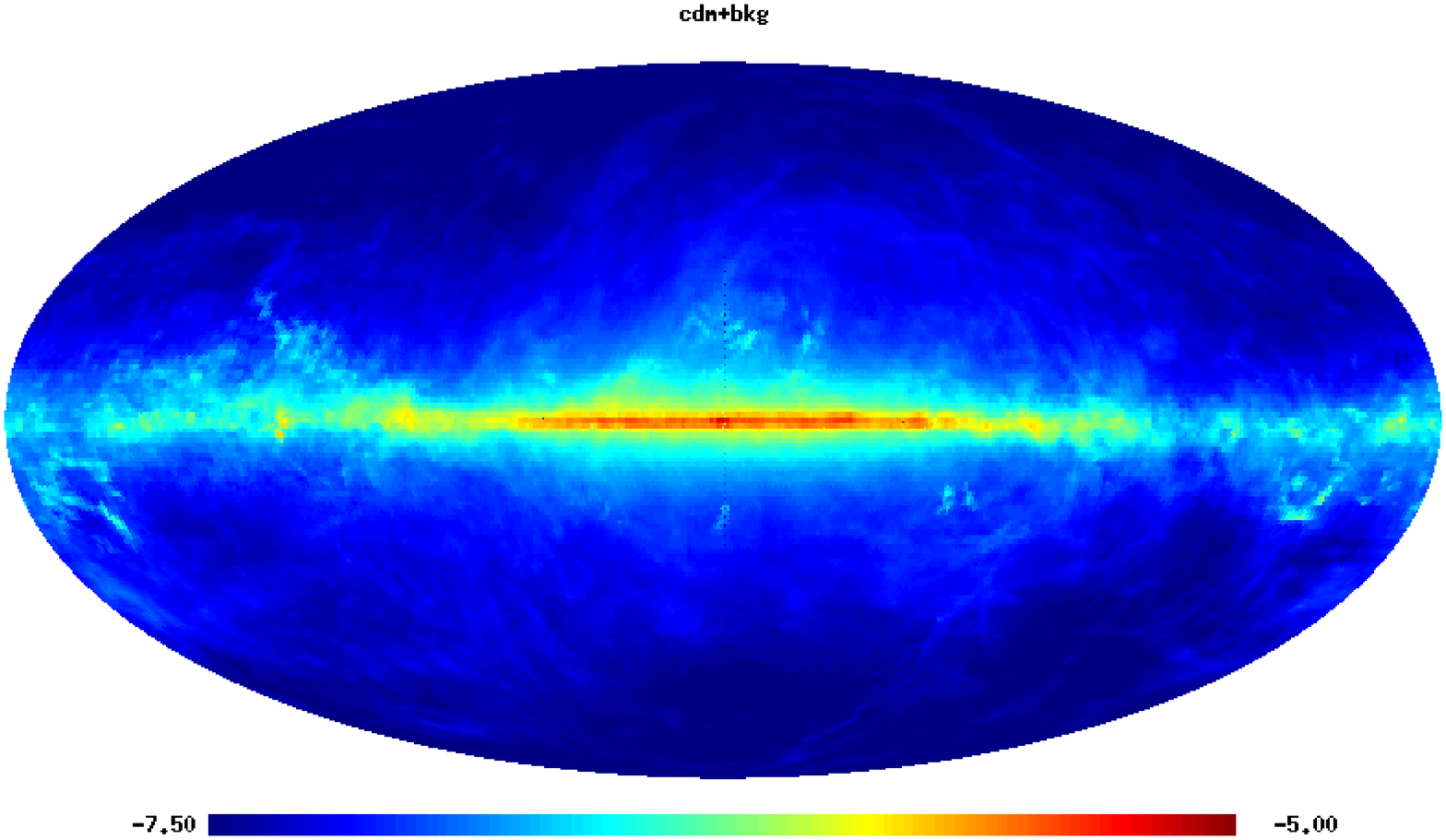}
\includegraphics[width=0.9\columnwidth]{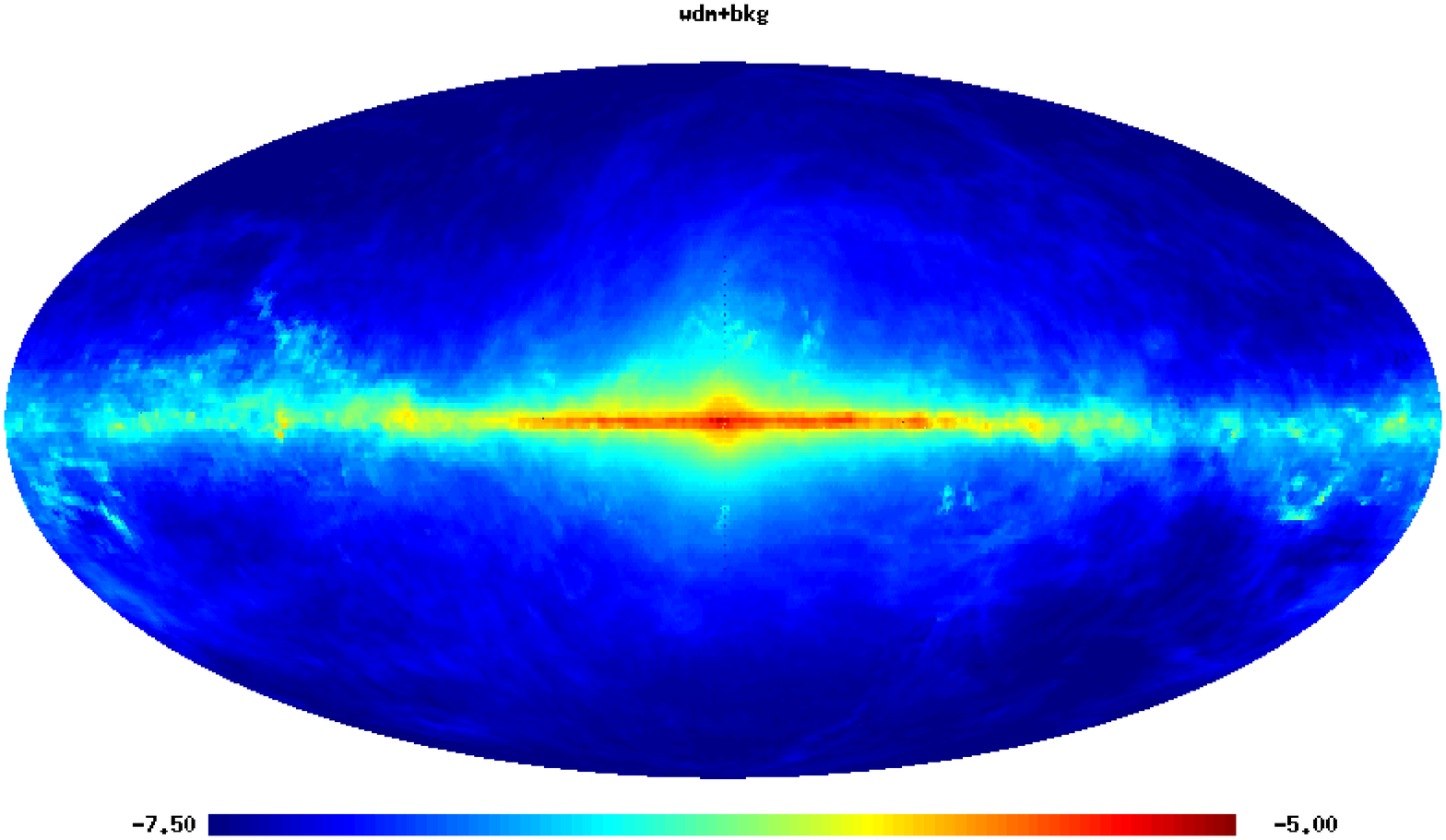}
\caption{Skymaps of the total $\gamma$-ray emission with background predicted
by GALPROP included, for energies $E>10$ GeV. The left panel is for the
cold WIMP case, and the right panel is for the warm WIMP case.
\label{fig:skymap_bkg}}
\end{figure*}

To discuss the detectability of DM we have to take the astrophysical
background of diffuse $\gamma$-rays into account. We use the
GALPROP\footnote{http://galprop.stanford.edu/}
\cite{1998ApJ...509..212S} code to calculate the Galactic diffuse
$\gamma$-ray background. The propagation parameters adopted are:
$D_0=6.59\times10^{28}$ cm$^2$s$^{-1}$, $\delta=0.30$, $v_A=39.2$
km s$^{-1}$, $z_h=3.9$ kpc, according to the fit to the B/C data
\cite{2011ApJ...729..106T}. The injection spectra of nuclei are adopted
as $\gamma_1^n=1.91$/$\gamma_2^n=2.40$ for rigidity below/above 10 GV,
which can basically reproduce the recent measurements of proton and
Helium spectra by PAMELA \cite{2011Sci...332...69A}, as shown in
\cite{2012PhRvD..85d3507L}. Note, however, this simple injection model
may not well describe the detailed hardening structures of the CR spectra
above several hundred GV, or the difference between proton and Helium
spectra (\cite{2009BRASP..73..564P,2010ApJ...714L..89A,
2011Sci...332...69A}). For CR electrons, the injection spectra are
$\gamma_1^e=1.50$/$\gamma_2^e=2.56$ for rigidity below/above $3.55$ GV as
derived according to the pure background fit to the newest $e^+e^-$ data
\cite{2012PhRvD..85d3507L}. Such a pure background component cannot
explain the $e^+e^-$ excesses revealed by several experiments
\cite{2009Natur.458..607A,2008Natur.456..362C,2008PhRvL.101z1104A,
2009A&A...508..561A,2009PhRvL.102r1101A}. As illustrated in
\cite{2010ApJ...720....9Z} the contribution to the total diffuse
$\gamma$-rays from the extra astrophysical sources of $e^+e^-$,
e.g., pulsars, is always negligible in all regions of the sky.
For the purpose of the current study, we think it is enough to employ
such a rough background model. Finally the extra-galactic
$\gamma$-ray background is adopted to be the Fermi measured results,
$\phi_{\rm EG}\approx 5.89\times 10^{-7}(E/{\rm GeV})^{-2.44}$ cm
$^{-2}$ s$^{-1}$ sr$^{-1}$ GeV$^{-1}$ \cite{2010PhRvL.104j1101A}.

We calculate the total $\gamma$-ray skymaps for the cold and warm WIMP
scenarios based on the two SUSY benchmark models given the previous
subsection. The total $\gamma$-ray skymaps above $10$ GeV of both the
astrophysical background and the DM contribution are shown in Fig.
\ref{fig:skymap_bkg}. The left panel is for cold WIMP and the right
panel is for warm WIMP respectively. The detectability of the DM
signal in presence of the astrophysical background will be discussed
in the followings two subsections.

\subsection{Gamma-rays from warm WIMP annihilation: Galactic center}

\begin{figure}[!htb]
\centering
\includegraphics[width=0.9\columnwidth]{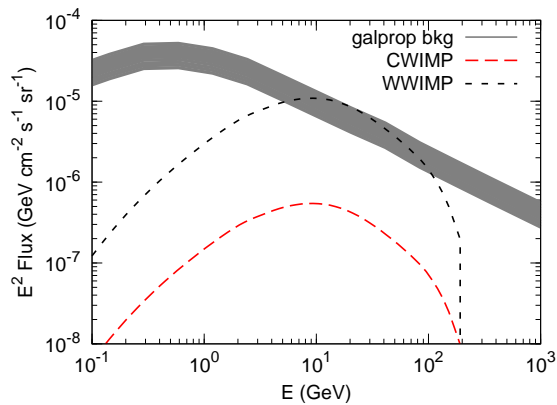}
\caption{Gamma-ray spectra in $20^{\circ}\times20^{\circ}$ region around
the Galactic center for cold and warm WIMP scenarios. Shaded region
represents the expected background of the GALPROP model (see the text).
\label{fig:gc}}
\end{figure}

Fig. \ref{fig:gc} shows the expected $\gamma$-ray spectra of the
WIMP annihilation in $20^{\circ}\times20^{\circ}$ region around
the Galactic center. For comparison we also plot the calculated
diffuse background described in Sec. IV. B. There are all-sky survey
data of diffuse $\gamma$-rays from Fermi, available from the Fermi
Science Support Center\footnote{http://fermi.gsfc.nasa.gov/ssc/}.
It was shown that the GALPROP model could reproduce the observational
data within a precision of factor $2$ \cite{2012ApJ...750....3A}.
Therefore here we simply employ the model results for comparison.
An uncertainty of $2$ times of the GALPROP background is represented
by the shaded region.

It is shown that for warm WIMP scenario we may expect larger flux of
$\gamma$-rays in the Galactic center, simply due to a larger annihilation
cross section of warm WIMPs. Up to now there is no clear evidence to
show the existence of signals from DM in the Fermi
data\footnote{See \cite{2011PhLB..697..412H} for a claim of DM signal
in the most central region of the Galactic center. However, the
background and possible point source contamination need to be carefully
studied.}. However, we may expect that the warm WIMP scenario could
have better detection perspective than the cold WIMP scenario.

\subsection{Gamma-rays from warm WIMP annihilation: subhalos}

\begin{figure}[!htb]
\centering
\includegraphics[width=0.9\columnwidth]{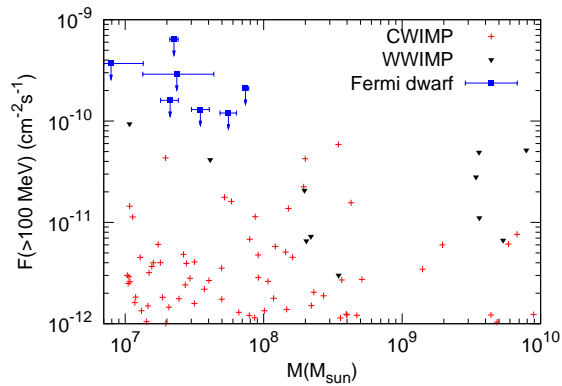}
\caption{Integral fluxes above $100$ MeV of the DM subhalos for the cold
and warm WIMP models. The arrows show the upper limits of dwarf galaxies
given by Fermi observations \cite{2010ApJ...712..147A}.
\label{fig:inte_flux}}
\end{figure}

Finally we discuss the detectability of DM subhalos. Fig. \ref{fig:inte_flux}
shows the integral fluxes above $100$ MeV of the DM subhalos for both the
cold and warm WIMP models. With a factor of $\sim 20$ times larger of the
cross section for warm WIMP scenario, we can see that the fluxes of the
most luminous subhalos in the two scenarios are comparable. Also shown
in Fig. \ref{fig:inte_flux} are the upper limits (for $80\%b\bar{b}+
20\%\tau^+\tau^-$ case) of dwarf galaxies derived through $11$-month
observations of Fermi \cite{2010ApJ...712..147A}. The upper limits are
in general higher than the model expected fluxes, which means the first
year Fermi data may not be able to probe the DM subhalos of both the
cold and warm WIMP models discussed here.

\begin{figure}[!htb]
\centering
\includegraphics[width=0.9\columnwidth]{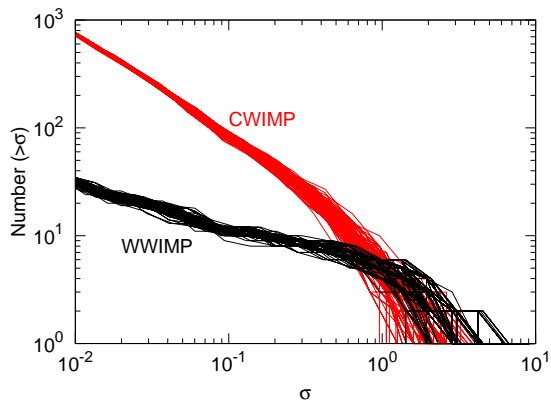}
\caption{The accumulative number of subhalos with significance higher
than $\sigma$, for energies $E>10$ GeV and $5$-yr exposure of Fermi-LAT
for the cold and warm WIMP scenarios.
\label{fig:Nsigma}}
\end{figure}

Fig. \ref{fig:Nsigma} gives the results of the accumulative number versus the
detection significance, defined as $\sigma=N_{\rm sig}/\sqrt{N_{\rm bkg}}$,
for $E>10$ GeV and $5$-yr exposure of Fermi. Here the emission from
subhalos within $\theta_{\rm half}$, angular radius containing half of
the annihilation luminosity, is taken into account. The sky range
to calculate the background number of events is adopted to be
$\max(\theta_{\rm half},\theta_{\rm res})$, where $\theta_{\rm res}
\approx0.1^{\circ}$, is the angular resolution of Fermi-LAT at $E>10$
GeV \cite{2009ApJ...697.1071A}.

Similar with Fig. \ref{fig:NJfactor} the number distribution for warm WIMP
is flatter than that for cold WIMP. This is a signature to distinguish these
two scenarios. It is interesting to note that the potential detectability
for warm WIMP might be a little bit better (for high $\sigma$ ones) than
that of cold WIMP, although the number of subhalos are significantly less.
This is because the allowed cross section for warm WIMP could be larger in
principle. However, it is generally difficult to detect the SUSY DM signals
from subhalos with the Fermi detector, either for the cold or the warm WIMP
scenarios.

\section{Conclusions}

Since more and more evidence shows that the DM tends to be ``warm''
instead of ``cold'' (e.g., \cite{2012EAS....53...97T}), it is necessary
to investigate the possible consequence on DM detections if it is indeed
warm. For the canonical light WDM particle like the sterile neutrino,
most of the present DM detection experiments will be useless. Alternatively
the non-thermally produced warm WIMP scenario \cite{1999JHEP...12..003J,
2001PhRvL..86..954L} might be interesting enough, for both the cosmological
structure formation and the detection of DM particles. The large
free-streaming of the DM may help to solve the problems of CDM scenario
at small scale, and the WIMP particles are able to be detected with most
of the experiments searching for DM.

Based on the high resolution numerical simulations of WDM structure
formation, we study the possible $\gamma$-ray signals from the
annihilation of warm WIMPs in the Milky Way. The {\it Aquarius} CDM
simulations are also employed to compare with the WDM results. We
investigate the expected skymaps of the DM annihilation, as well as
the statistical properties of the subhalos. The detectability with
Fermi telescope is also discussed for two benchmark SUSY models of
warm and cold WIMP scenarios respectively. Unfortunately we find
that the detectability of the warm WIMPs with current $\gamma$-ray
experiments is very poor. Nevertheless, it is interesting to investigate
the theoretically expected signatures of the $\gamma$-rays from warm
WIMP annihilation, in case that they might be detected in future.

The major conclusions of this work can be summarized as follows.

\begin{itemize}

\item Due to a suppression of structure formation in WDM scenario,
  subhalo is much less abundant in WDM scenario, resulting in a flatter
  accumulative number distribution of $J$-factor and a different $N(>J)$
  vs. $J$ relation between warm and cold WIMP models.

\item We find it is difficult to detect the subhalos with Fermi
  telescope, both for cold and warm WIMP scenarios. It is found that
  the detectablity of warm WIMP could in principle be better than cold
  WIMP, because a moderately larger annihilation cross section is
  allowed for warm WIMP scenario, with a non-thermal production
  mechanism \cite{2009PhRvD..80j3502B}.

\item For indirect WIMP search strategy, the Galactic center would
  likely be prior to dwarf galaxies if DM is made of warm WIMPs. For
  cold WIMPs the $\gamma$-ray emission due to dark matter annihilation
  from the Galactic center is polluted by the high background and the
  subhalos have been believed to be better targets for DM indirect
  searches. In the warm WIMP case, however, the emission from the
  Galactic center could be enhanced due to a larger cross section,
  while the emission from dwarf galaxies is not as significantly
  enhanced because of the decrease of the central DM density and
  concentration. For our benchmark models, the signal of the warm WIMP
  annihilation from the Galactic center will be $\sim20$ times stronger
  than that of cold WIMPs, while it is comparable for subhalos. This
  might lead to a different detection strategy in case that WIMP is warm.

\end{itemize}

\section*{Acknowledgements}

We thank Paolo Gondolo, Shi Shao and Charling Tao for useful discussion,
and the anonymous referee for helpful comments.
This work is supported by National Natural Science Foundation of China
under grant Nos. 11075169, 11105155, 11105157, 11033005, 10975142,
10973018, 11133003, the 973 project under grant Nos. 2010CB833000,
2009CB24901, and Chinese Academy of Sciences under grant No. KJCX2-EW-W01.

\begin{appendix}

\section{Relic density of non-thermal DM from cosmic string decay}

Here we briefly discuss the non-thermal DM density from cosmic string 
decay. We assume the correlation length scale of the string is $\xi(t)$ 
in the friction dominated epoch. It can be given by $\xi(t)=\xi(t_c)
(t/t_c)^{3/2}$ \cite{1999PhLB..445..323B}, where the initial length
$\xi(t_c)\sim \lambda^{-1} \eta^{-1}$, $\lambda$ is the scalar
self-quartic coupling. The production of cosmic string loops induce
the energy lose of long strings. The number density of loops created
by long strings can be evaluated by \cite{1995RPPh...58..477H,
1994IJMPA...9.2117B}
\begin{equation}
\frac{{\rm d}n}{{\rm d}t}=\nu \xi^{-4} \frac{{\rm d}\xi}{{\rm d}t},
\end{equation}
where $\nu$ is a constant of order 1. We assume each loop contributes
$N$ DM particles. 

Here we only consider the non-thermal DM particles from the decay of 
loops below the temperature $T_\chi$ (the corresponding time is $t_chi$).
For $m_{\chi}\sim 100$ GeV, $T_\chi\sim$GeV. Then we can get the DM number 
density by integrating the red-shifted cosmic string loop number density
\begin{equation}
n^{\rm NT}_{\chi}(t_0)=N \nu \int^{\xi_0}_{\xi(t_F)} \left(\frac{t}
{t_0}\right)^\frac{3}{2} \xi^{-4} {\rm d}\xi,
\end{equation}
where $t_F$ is the time at which cosmic string loops which are decaying
at $t_\chi$ form. Since the loop density decreases sharply with time,
we can see the DM density is mainly contributed by loops which decay
right after $t_\chi$. It means the most of non-thermal DM particles are
created at $t_\chi$ instantaneously. 

According to the average radius of loop (formed at $t_F$) $R(t_F) \sim
\lambda^{\frac{1}{2}} g^{* \frac{3}{4}}_{t_F} G \mu
M^\frac{1}{2}_{pl}t^\frac{3}{2}_F$, and the loop shrink rate
${\rm d}R/{\rm d}t=-\Gamma G \mu$ ($\Gamma$ is a constant $\sim 10-20$)
\cite{1995RPPh...58..477H,1994IJMPA...9.2117B},
we find $t_F \sim \lambda^{-\frac{1}{3}} g^{* -\frac{1}{2}}_{t_F}
\Gamma^{\frac{2}{3}} M^{-\frac{1}{3}}_{pl}t^\frac{2}{3}_\chi $ .
Then the reduced number density of non-thermal DM particles from decays
of cosmic string loops can be derived as \cite{1999JHEP...12..003J,
2009PhRvD..80j3502B}
\begin{equation}
Y^{\rm NT}_{\chi}=\frac{6.75}{\pi} N \nu \lambda^{\frac{3}{2}}
\Gamma^{-2} g^{* \frac{3}{2}}_{t_c} g^*_{t_\chi} g^{* -\frac{5}{2}}_{t_F}
M^2_{pl} \frac{T^4_\chi}{T^6_c},
\end{equation}
where $Y_{\chi}$ is defined as $Y_{\chi} = n_{\chi}/s$,
$s=2\pi^2g_{*}T^3/45$ is the entropy density, $g^{*}$ is effective
degrees of freedom at the corresponding time. The DM relic density is
related to $Y$ by $\Omega h^2 = 2.82 \times 10^8 Y_{\chi} (m_{\chi}/
{\rm GeV})$. If the DM is dominated by non-thermal production, we can
get the corrected DM relic density $\Omega h^2 \sim 0.11$ easily by
choosing $\nu, \lambda \sim 1$, $\Gamma \sim 10$, $g^{*} \sim 100$,
$m_{\chi} \sim O(10^2)$ GeV and $T_c \sim O(10^9)$ GeV
\cite{1999JHEP...12..003J,2009PhRvD..80j3502B}.

\section{DM distribution from {\it Aquarius} simulation}

Here we give the statistical results used for the extrapolation of the
unresolved subhalos, based on the resolved subhalos from {\it Aquarius}
CDM simulations. More results can be found in Refs.
\cite{2008MNRAS.391.1685S,2011MNRAS.410.2309G}.

\begin{figure*}[!htb]
\centering
\includegraphics[width=0.9\columnwidth]{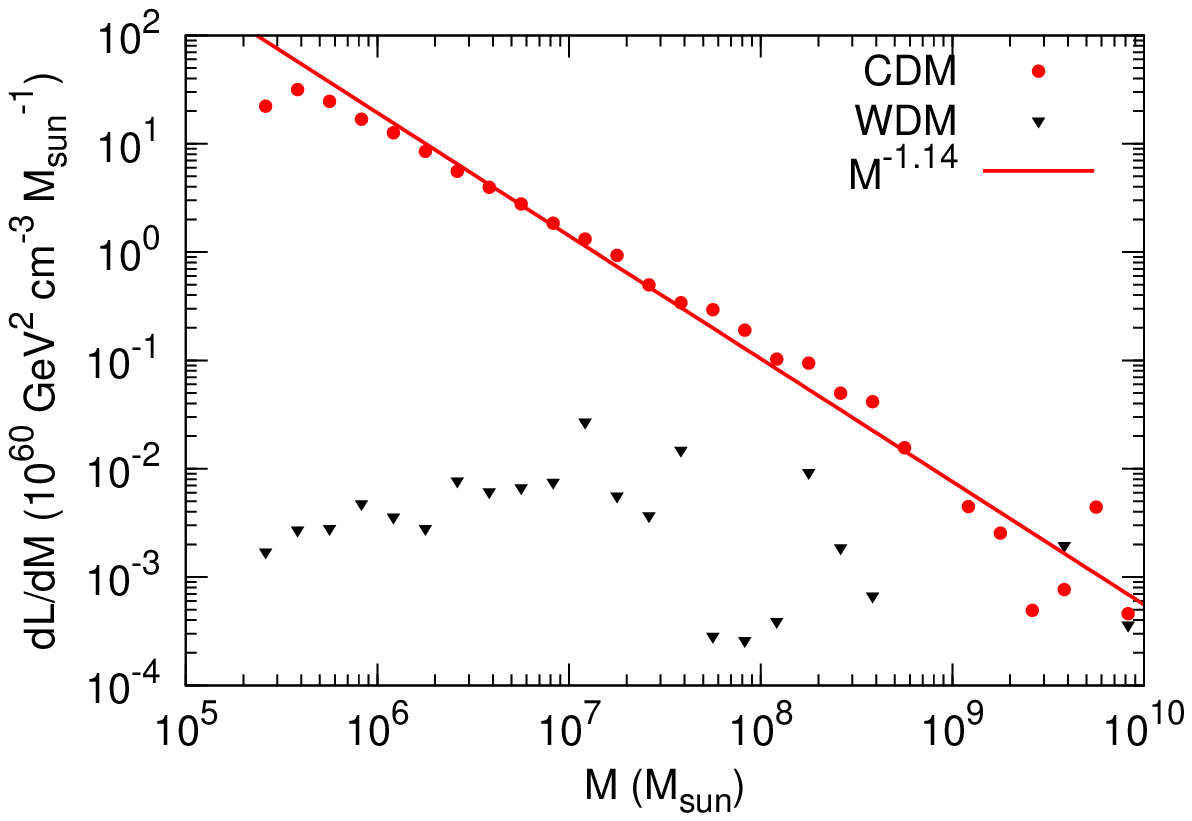}
\includegraphics[width=0.9\columnwidth]{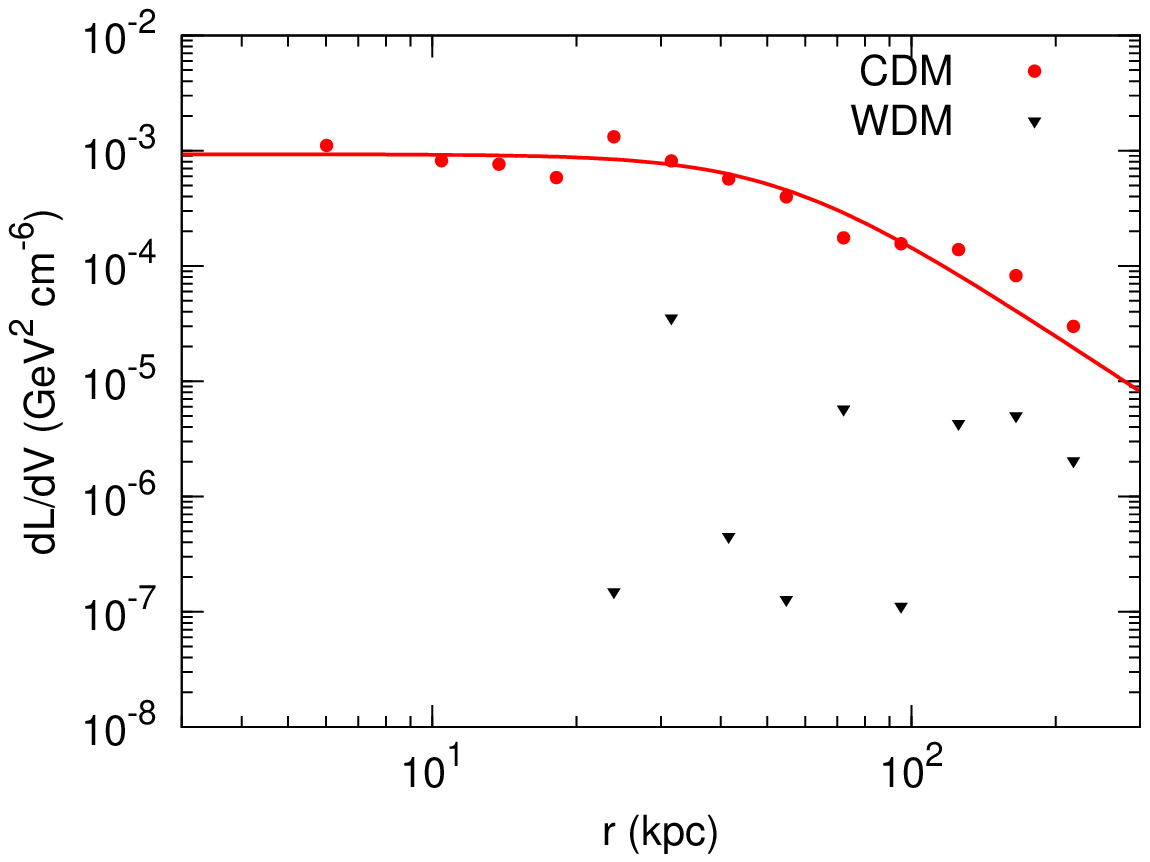}
\caption{Differential luminosity-mass relation ${\rm d}L/{\rm d}M$ (left)
and spatial density of luminosity (right) for subhalos. Red circles are
for Aq-A-2 CDM simulation, and black triangles are for Aq-AW-2 WDM
simulation. Solid lines are the fitting results for CDM.
\label{fig:dLdMdV}}
\end{figure*}

We bin the luminosities $L_i$ of the subhalos with respect to mass and
radius. The left panel of Fig. \ref{fig:dLdMdV} shows the differential
distribution of luminosity versus subhalo mass, ${\rm d}L/{\rm d}M$,
and the right panel shows the spatial distribution of the luminosity,
${\rm d}L/{\rm d}V$. When doing this analysis we assume that
${\rm d}L/{\rm d}M$ is independent with the spatial distribution
${\rm d}L/{\rm d}V$ \cite{2008Natur.456...73S}, so that we can use all
the subhalos to derive both ${\rm d}L/{\rm d}M$ and ${\rm d}L/{\rm d}V$.
The results for WDM  are also shown in Fig. \ref{fig:dLdMdV} for comparison.

\begin{figure*}[!htb]
\centering
\includegraphics[width=0.9\columnwidth]{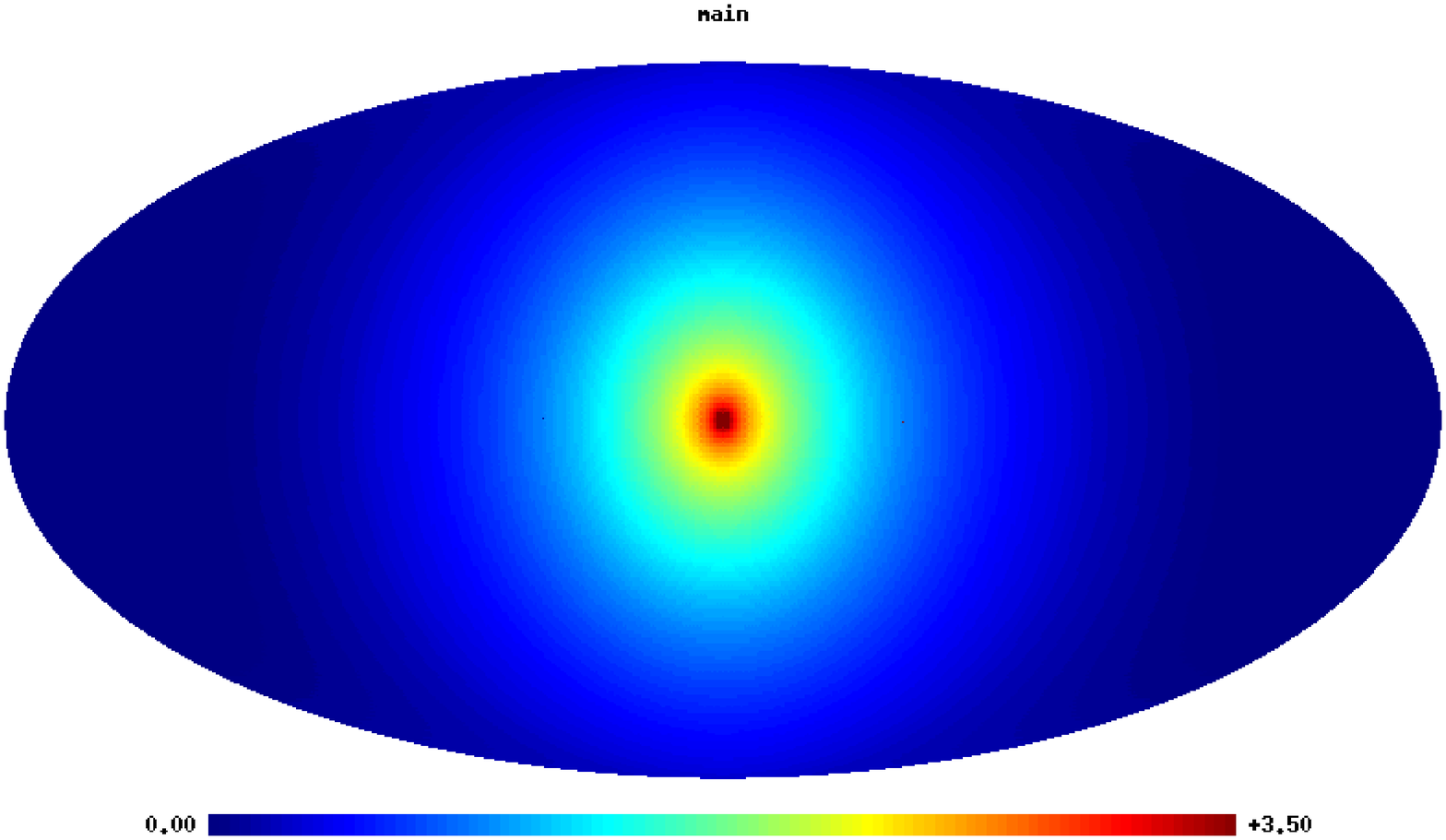}
\includegraphics[width=0.9\columnwidth]{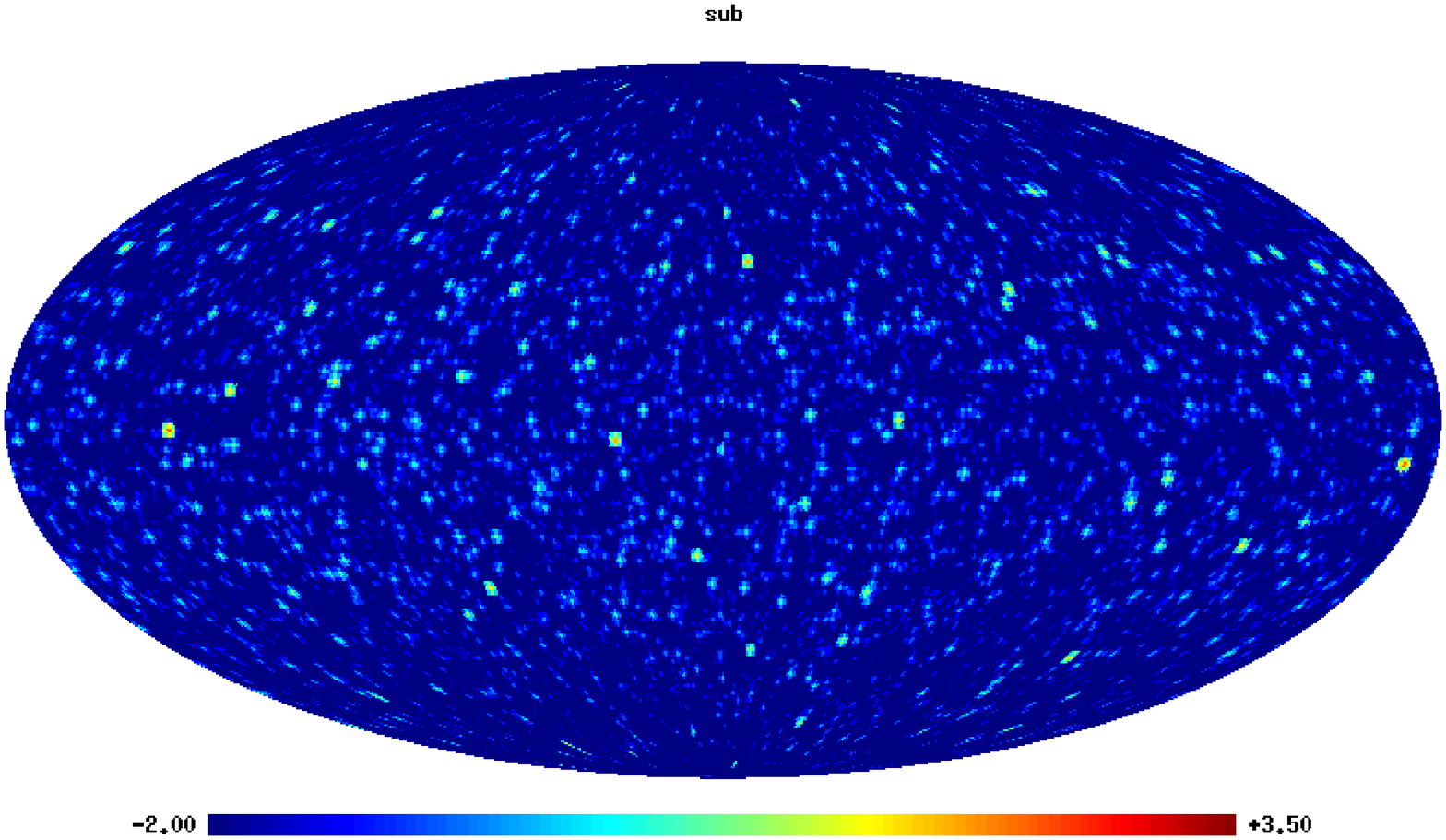}
\includegraphics[width=0.9\columnwidth]{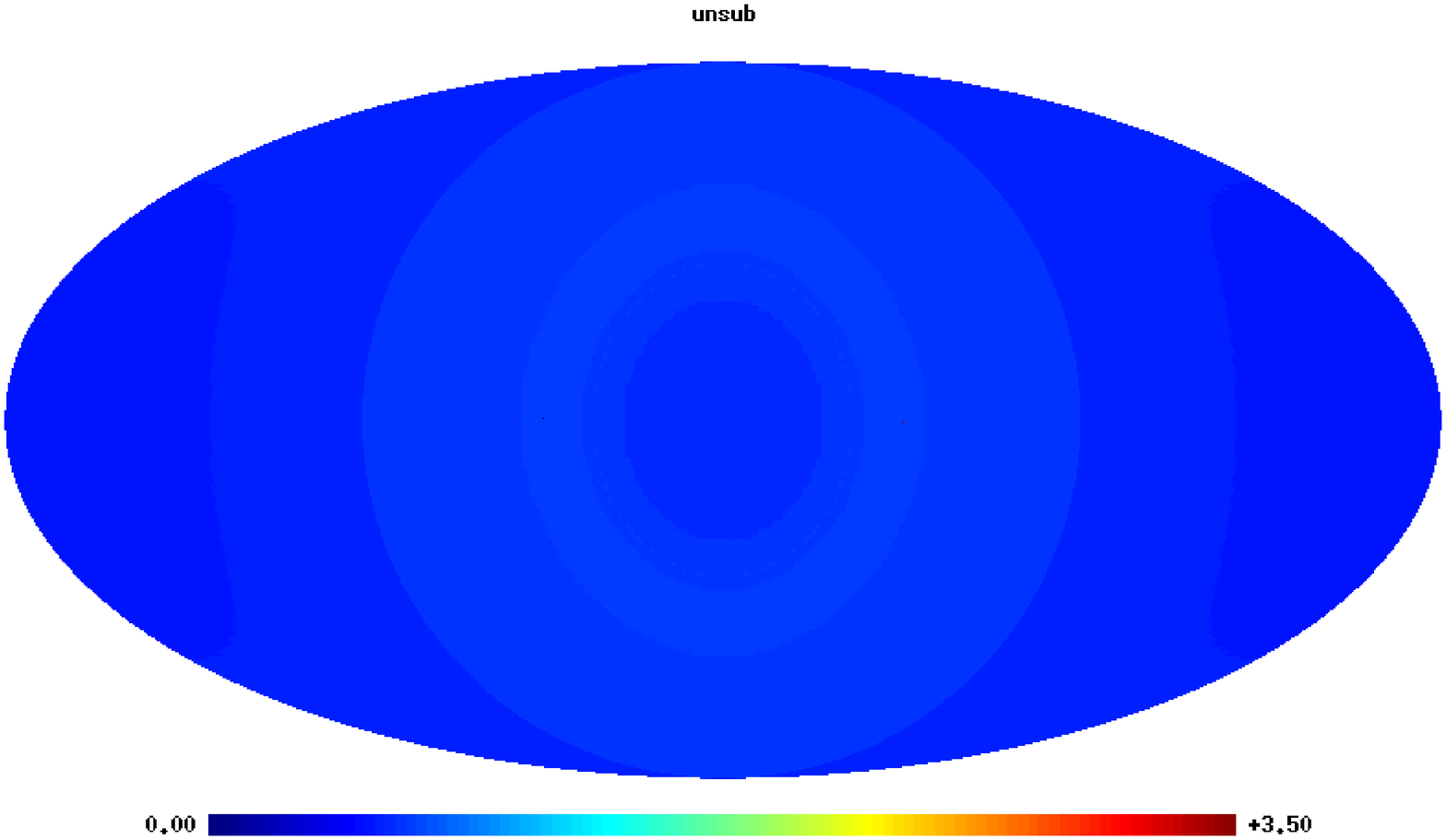}
\includegraphics[width=0.9\columnwidth]{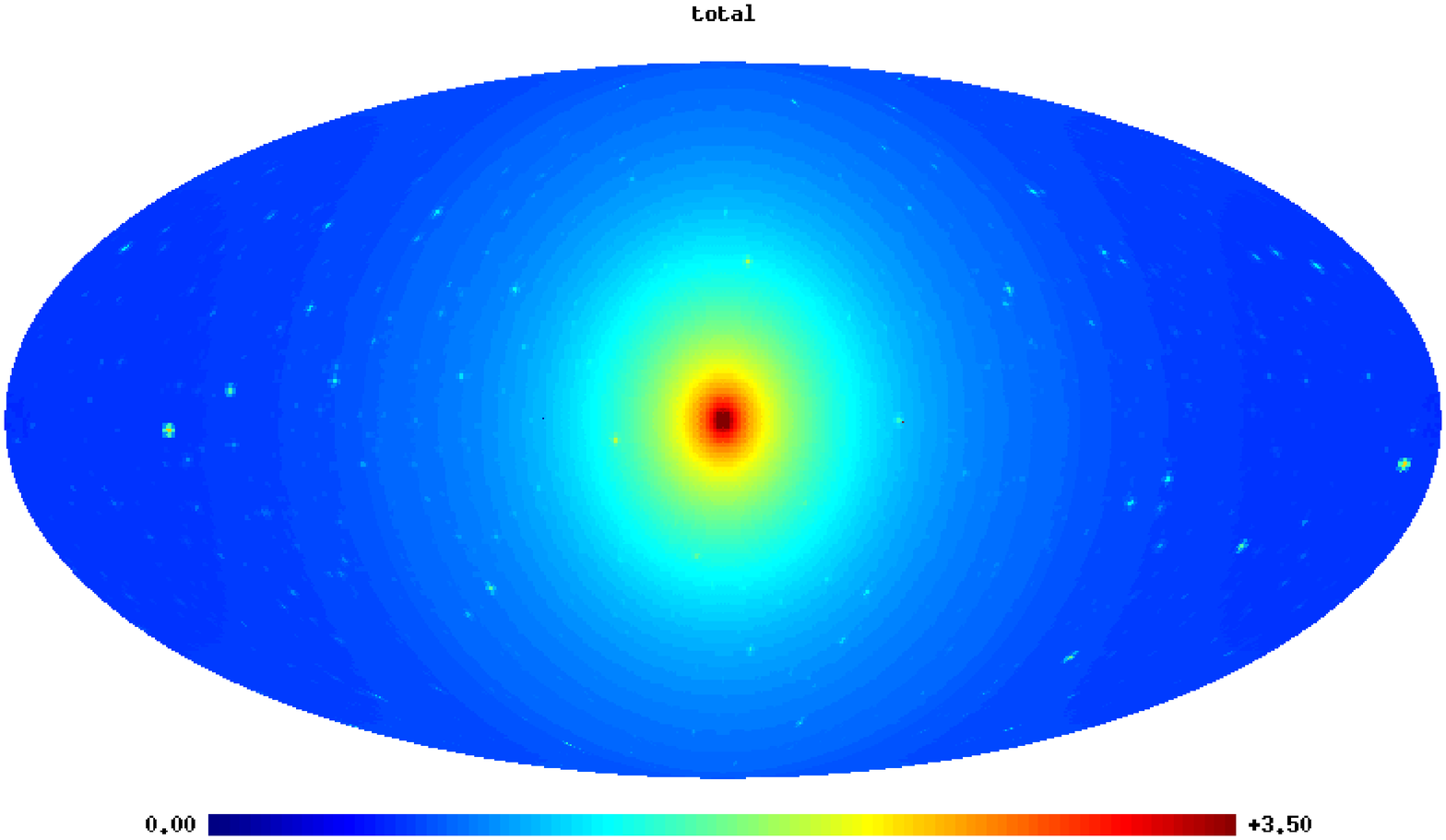}
\caption{Skymaps of the $J$-factors of the smooth halo (top-left),
resolved subhalos (top-right), unresolved subhalos (bottom-left)
and the total contribution (bottom-right) for CDM.
\label{fig:skymap_cdm}}
\end{figure*}

The luminosity-mass relation ${\rm d}L/{\rm d}M$ can be fitted with a
power-law function
\begin{equation}
{\rm d}L/{\rm d}M\propto M^{-1.14}.
\label{dLdM}
\end{equation}
We can infer the cumulative luminosity distribution as $L(>M_{\rm th})
\propto M_{\rm th}^{-0.14}-M_{\rm max}^{-0.14}$. For $M_{\rm th}\ll
M_{\rm max}$ we have $L(>M_{\rm th})\propto M_{\rm th}^{-0.14}$. Note
that this result is flatter than the mass dependence of the cumulative
luminosity derived in \cite{2008Natur.456...73S} ($\propto M_{\rm th}^
{-0.226}$). This may be due to the threshold effect when $M_{\rm th}$
is close to $M_{\rm max}\approx 10^{10}$ M$_{\odot}$.
For the spatial distribution of the luminosity ${\rm d}L/{\rm d}V$
we use an iso-thermal $\beta$ function
\begin{equation}
{\rm d}L/{\rm d}V\propto \frac{1}{\left[1+(r/r_c)^{\beta}\right]}
\end{equation}
to fit the simulation results. The fitting parameters are $r_c\approx
54$ kpc and $\beta\approx 2.76$.

The unresolved subhalos in the CDM simulation is derived according to
the fitting results of ${\rm d}L/{\rm d}M$ and ${\rm d}L/{\rm d}V$.
The masses of unresolved subhalos are assumed to extend to $M_{\rm min}
\approx 10^{-6}$ M$_{\odot}$ from $M_{\rm res}\approx 3\times10^5$
M$_{\odot}$.

The $J$-factor for unresolve subhalos is
\begin{equation}
J_{\rm sub}^{\rm un}(\psi)=\frac{1}{\rho_{\odot}^2R_{\odot}}\int_{LOS}
\left(\int_{M_{\rm min}}^{\rm M_{\rm res}}\frac{{\rm d}^2L}{{\rm d}M
{\rm d}V}{\rm d}M\right){\rm d}l.
\label{jpsi_sub_un}
\end{equation}
Fig. \ref{fig:skymap_cdm} shows the skymaps of $J$-factors of the smooth
halo (top-left), resolved subhalos (top-right), unresolved subhalos
(bottom-left) and the total contribution (bottom-right) for CDM. This
figure is a reproduction of the result given in Ref.
\cite{2008Natur.456...73S}.

\end{appendix}

\bibliography{/home/yuanq/work/cygnus/tex/refs}
\bibliographystyle{apsrev}

\end{document}